\documentclass[aps, pra, twocolumn,showpacs,preprintnumbers,amsmath,amssymb]{revtex4}

\newcommand{\ket}[1]{\mbox{$ | #1 \rangle $}}
\newcommand{\bra}[1]{\mbox{$ \langle #1 | $}}
\usepackage[dvips]{graphicx}

\begin{document}

\preprint{}

\title{Quantum secure direct communication based on order rearrangement of single photons}

\author{Jian Wang}

 \email{jwang@nudt.edu.cn}

\affiliation{School of Electronic Science and Engineering,
\\National University of Defense Technology, Changsha, 410073, China }
\author{Quan Zhang}
\affiliation{School of Electronic Science and Engineering,
\\National University of Defense Technology, Changsha, 410073, China }
\author{Chao-jing Tang}
\affiliation{School of Electronic Science and Engineering,
\\National University of Defense Technology, Changsha, 410073, China }


\begin{abstract}
Based on the ideal of order rearrangement and block transmission of
photons, we present a quantum secure direct communication scheme
using single photons. The security of the present scheme is ensured
by quantum no-cloning theory and the secret transmitting order of
photons. The present scheme is efficient in that all of the
polarized photons are used to transmit the sender's secret message
except those chosen for eavesdropping check. We also generalize this
scheme to a multiparty controlled quantum secret direct
communication scheme which the sender's secret message can only be
recovered by the receiver under the permission of all the
controllers.
\end{abstract}

\maketitle


%
%
Quantum cryptography has been one of the most promising applications
of quantum information science. It utilizes quantum effects to
provide unconditionally secure information exchange. Since the first
QKD protocol was proposed by Benneett and Brassard in 1984
\cite{bb84}, quantum key distribution (QKD) which provides
unconditionally secure key exchange has progressed quickly. In
recent years, a good many of other quantum cryphtography schemes
have also been proposed and pursued, such as quantum secret sharing
(QSS)\cite{hbb99,kki99,zhang,gg03,zlm05,xldp04}, quantum secure
direct communication (QSDC)
\cite{beige,Bostrom,Deng,denglong,cai1,cai4,jwang1,jwang2,cw1,cw2,tg,zjz}.
QSS is the generalization of classical secret sharing to quantum
scenario and can share both classical and quantum messages among
sharers. QSDC's object is to transmit the secret message directly
without first establishing a key to encrypt them, which is different
to QKD. QSDC can be used in some special environments which has been
shown by Bostr\"{o}em and Deng et al.\cite{Bostrom,Deng}. Many
researches have been carried out in QSDC. We can divide these works
into two types, one utilizes single photons \cite{denglong,cai1},
the other utilizes entangled state
\cite{Bostrom,Deng,cai1,cai4,jwang1,jwang2,cw1,cw2,tg,zjz}. Deng et
al. proposed a QSDC scheme using batches of single photons which
serves as one-time pad \cite{denglong}. Cai et al. presented a
deterministic secure direct communication scheme using single qubit
in a mixed state \cite{cai1}. The QSDC scheme using entanglement
state is certainly the mainstream. Bostr\"{o}m and Felbinger
proposed a "Ping-Pong" QSDC protocol which is quasi-secure for
secure direct communication if perfect quantum channel is used
\cite{Bostrom}. Cai et al. pointed out that the "Ping-Pong" Protocol
is vulnerable to denial of service attack or joint horse attack with
invisible photon \cite{cai2,cai3}. They also presented an improved
protocol which doubled the capacity of the "Ping-Pong" protocol
\cite{cai4}. Deng et al. put forward a two-step QSDC protocol using
Einstein-Podolsky-Rosen (EPR) pairs \cite{Deng}. We presented a QSDC
scheme using EPR pairs and teleportation \cite{jwang1} and a
multiparty controlled QSDC scheme using Greenberger-Horne-Zeilinger
states \cite{jwang2}.

In Ref. \cite{deng-core}, Deng et al. utilize controlled order
rearrangement encryption to realize a QKD scheme. In their scheme,
the communication parties share a control key used to control the
order rearrangement operation. Very recently, A. D. Zhu et al.
proposed a QSDC scheme based on secret transmitting order of
entangled particles \cite{zhu}. The security of their scheme is
based on entanglement and the secret transmitting order of
particles. In this Letters, we present a QSDC scheme using single
photons based on the ideal of the order rearrangement and qubit
transmission in batches \cite{deng-core,zhu,long}. The initial state
of the transmitting photon is prepared randomly in one of the four
states belonging to two conjugate basis, which is similar to the
BB84 QKD protocol. The security of the scheme is based on quantum
no-cloning theory and the secret transmitting order of single
photons. All of the single photons are used to generate secret
message except those used for eavesdropping check. It is not
necessary for the communication parties to choose a random measuring
basis for eavesdropping check. Compared with schemes using EPR
pairs, this scheme is more realizable. We also generalize this
scheme to a multiparty controlled quantum secret direct
communication (MCQSDC) scheme. In the MCQSDC scheme, the sender's
secret message is transmitted directly to the receiver and can only
be reconstructed by the receiver with the permission of all the
controllers. We also discuss the security of the two schemes, which
is unconditionally secure.


Here We first describe the details of our QSDC scheme using single
photons. Suppose the sender Bob wants to transmit his secret message
directly to the receiver Alice.

(S1) Alice prepares $N$ single photons each of which is randomly in
one of the following states
\begin{eqnarray}
& &\ket{H}=\ket{0},\nonumber\\
& &\ket{V}=\ket{1},\nonumber\\
& &\ket{+}=\frac{1}{\sqrt{2}}(\ket{0}+\ket{1}),\nonumber\\
& &\ket{-}=\frac{1}{\sqrt{2}}(\ket{0}-\ket{1}).
\end{eqnarray}
She then send the $N$ photons $[P_1,P_2,\cdots,P_n]$, called
$P$-sequence to Bob.

(S2) Bob selects randomly a sufficiently large subset from the
$P$-sequence for eavesdropping check, called checking sequence
($C$-sequence). The remaining photons of the $P$-sequence form a
message sequence ($M$-sequence). Bob performs randomly one of the
two unitary operations
\begin{eqnarray}
& &I=\ket{0}\bra{0}+\ket{1}\bra{1},\nonumber\\
& &U=i\sigma_y=\ket{0}\bra{1}-\ket{1}\bra{0}.
\end{eqnarray}
on each of the photons in the $C$-sequence. He also encode his
secret message on the $M$-sequence by performing one of the unitary
operations $I$ and $U$, according to his secret message. If his
secret message is ``0'' (``1''), Bob performs operation $I$ ($U$).
The operation $U$ flips the state in both $Z$-basis
(\ket{0},\ket{1}) and $X$-basis (\ket{+},\ket{-}), as
\begin{eqnarray}
& &U\ket{0}=-\ket{1}, U\ket{1}=\ket{0},\nonumber\\
& &U\ket{+}=\ket{-}, U\ket{-}=-\ket{+}.
\end{eqnarray}

(S3) Bob disturbs the order of the photons in the $P$-sequence and
generates a rearranged photon sequence, called $P'$-sequence [$P_1',
P_2',\cdots,P_n'$]. He then sends the $P'$-sequence to Alice. The
order of $P'$-sequence is completely secret to others but Bob
himself, which ensures the security of the present scheme.

(S4) Alice tells Bob she has received the $P'$-sequence. After
hearing from Alice, Bob announces the position of the $C$-sequence
and the secret rearranged order in it. He also publishes his
corresponding operations on the photons in the $C$-sequence.

(S5) Alice has the initial state information and the position of
each checking photons. She then performs von Neumann measurement on
each of the checking photons. If the initial state of the checking
photon is \ket{H} or \ket{V} (\ket{+} or \ket{-}), Alice performs
$Z$-basis ($X$-basis) measurement on it. Alice has the initial
states information of the checking photons, Bob's operation
information on the checking photons and her measurement results on
them. She can then evaluate the error rate of the transmission of
the $P$-sequence. If the error rate exceeds the threshold, they
abort the scheme. Otherwise, they continue to the next step.

(S6) Bob publishes the secret order of the $M$-sequence. According
to the initial states information of the $M$ sequence, Alice
performs $Z$-basis or $X$-basis measurement on the $M$-sequence. She
can then obtain Bob's secret message.

We now discuss the unconditional security of the present scheme. The
security of the scheme is based on quantum no-cloning theory and the
secret transmitting order of the photons. Quantum no-cloning theory
ensures that an eavesdropper, Eve cannot make certain the initial
states of the transmitting photons prepared by Alice, which is
similar to the BB84 QKD protocol. The difference between the BB84
QKD protocol and the present scheme is that the communication
parties perform $Z$-basis or $X$-basis measurement randomly for
preventing eavesdropping in the former, but the order rearrangement
is used to prevent Eve from obtaining the sender's secret message in
the latter. Suppose Eve intercepts the $P$-sequence and resends
another photon sequence prepared by Eve to Bob. After Bob has
performed his operations on the photon sequence, he then sends it to
Alice. Eve can also intercepts this photon sequence. However, Eve
cannot obtain Bob's operation information because Bob disturbs the
order of the photon sequence and Eve's attack will be detected
during the eavesdropping check. Without the correct order of the
photon sequence, Eve can only obtain a batch of meaningless data.
Obviously, the present scheme is also safe against collective attack
due to the secret order of rearranged photon sequence. As we
described above, the present scheme is unconditionally secure.

We then generalize this QSDC scheme to a MCQSDC scheme. Suppose the
sender Bob wants to transmit his secret message directly to the
receiver Alice under the control of the controllers Charlie,
Dick,$\cdots$, York and Zach.

(S1$'$) Alice prepares a batch of N single photons randomly in one
of the four states \ket{H}, \ket{V}, \ket{+}, \ket{-} and sends this
batch of photons to Charlie.

(S2$'$) Charlie performs randomly one of the three unitary
operations $I$, $U$, $H$ on each photon, where
\begin{eqnarray}
H=\frac{1}{\sqrt{2}}(\ket{0}\bra{0}-\ket{1}\bra{1}+\ket{0}\bra{1}+\ket{1}\bra{0})
\end{eqnarray}
is a Hadamada operation. $H$ can realize the transformation between
$Z$-basis and $X$-basis,
\begin{eqnarray}
& &H\ket{0}=\ket{+}, H\ket{1}=\ket{-},\nonumber\\
& &H\ket{+}=\ket{H}, H\ket{-}=\ket{V}.
\end{eqnarray}
He then sends the $N$ photons to the next controller, say Dick. Dick
and the remaining controllers repeat the similar operations as
Charlie until Zach finishes his operations on the $N$ photons. Zach
then sends the $N$ photons to Bob.

(S3$'$) Similar to (S2), Bob encodes his random message and secret
message on the $C$-sequence and $M$-sequence, respectively. He also
disturbs the order of the $P$-sequence and send the rearranged
$P$-sequence to Alice.

(S4$'$) After hearing from Alice, Bob announces the $C$-sequence and
the secret order in it. He then let Alice publish the initial states
of the sampling photons. To prevent Alice's intercept-resend attack,
for each of the sampling photons Bob selects randomly a controller
to announce his or her $H$ operation information on the sampling
photon firstly and then the others do in turn. That is to say, the
controllers do not publish their $I$ or $U$ operation information
but only publish their $H$ operation information on the sampling
photons. According to these information, Alice can measure the
sampling photons in a correct measuring basis. She tells Bob his
measurement results. Bob chooses randomly a controller to publish
his or her $I$, $U$ operation information on each of the sampling
photons firstly and then the others do one by one. Thus Bob can
determine the error rate of the transmission of the $P$ sequence. If
he confirms there is no eavesdropping, the process is continued.
Otherwise, the process is stopped.

(S5$'$) Bob publishes the secret order of the $M$-sequence. If the
controllers permit Alice to reconstruct Bob's secret message, they
tell Alice their operation information. Thus Alice can obtain Bob's
secret message under the permission of the controllers Charlie,
$\cdots$, York, Zach.

The $H$ operation performed by the controllers is very important for
the security of the scheme. The nice feature of the $H$ operation
which can realize the transformation between $Z$-basis and $X$-basis
can prevent Eve or a dishonest controller from obtaining the control
information of the controllers. If the controllers only performs $I$
or $i\sigma_y$ operations on the $P$ sequence, Eve or a dishonest
Alice can obtain the control information by taking intercept-resend
attack. Eve intercepts the $P$-sequence and resends a fake photon
sequence to the controller. After the controller has performed his
or her operations on the $P$-sequence and sent it to the next
controller, Eve can also intercept the photon sequence and then
obtain the controller's information by measuring it. Bob firstly let
the controllers publish their $H$ operation information and then
Alice can choose a correct measuring basis. Only after Alice has
published her measurement results could the controllers announce
their $I$, $U$ operation information. It can prevent Alice from
obtaining Bob's secret message without the control of the
controllers. If the controllers publish all of their operation
information firstly, Alice can break the control of the controllers
by taking intercept-resend attack. In this attack, she sends the
$P$-sequence to Bob directly and a fake photon sequence to Charlie.
Certainly, she should intercept the photon sequence which Zach sends
to Bob. With their controller's operation information, during the
eavesdropping check Alice can successfully deceive Bob and then
obtain his secret message without the permission of the controllers.
Bob chooses randomly a controller to publish his or her operation
information on each of the sampling photons firstly and then the
others do one by one during the eavesdropping check. It ensures each
controller can really act as a controller. Suppose Bob let Charlie,
Dick, $\cdots$, York publish their operation information firstly and
Zach do finally. In other words, Bob does not select randomly a
controller to publish his or her information. Alice can then
collaborate with Zach to acquire Bob's secret message without the
control of other controllers. Alice sends the $P$-sequence to Zach
directly and a fake sequence to Charlie. Zach resends the
$P$-sequence to Bob without doing any operation. Zach can know what
operation information he should publish according to the operation
information of the controllers Charlie, Dick, $\cdots$, York. The
attack of Alice and Zach will not be detected by Bob during the
eavesdropping check. On the basis of the above analysis, the present
MCQSDC scheme is secure.

So far we have presented a QSDC scheme and a MCQSDC scheme using
single photons. Quantum no-cloning theory and the secret
transmitting order of photons ensure the security of these schemes.
In these schemes, all of the polarized photons are used to transmit
the sender's secret message directly to the receiver except those
chosen for checking eavesdropping. During the process of the scheme,
it only needs once eavesdropping check. Compared with the schemes
using entangled state, these schemes are practical within the
present technology.



\begin{acknowledgments}
This work is supported by the National Natural Science Foundation of
China under Grant No. 60472032.
\end{acknowledgments}

%
%

%
%

\begin{thebibliography}{99}
\bibitem{bb84} C. H. Bennett and G. Brassard, \textit{in Proceedings of IEEE
international Conference on Computers, Systems and signal
Processing, Bangalore, India} (IEEE, New York), pp. 175 - 179
(1984).
\bibitem{hbb99} M. Hillery, V. Buz\v{e}k, and A. Berthiaume, Phys. Rev. A \textbf{59}, 1829 (1999).
\bibitem{kki99} A. Karlsson, M. Koashi, and N. Imoto, Phys. Rev. A \textbf{59}, 162 (1999).
\bibitem{zhang} Z. J. Zhang, Phys. Lett. A \textbf{342}, 60 (2005).
\bibitem{gg03} G. P. Guo and G. C. Guo, Phys. Lett. A \textbf{310}, 247 (2003).
\bibitem{zlm05} Zhan-jun Zhang, Yong Li, and Zhong-xiao Man, Phys. Rev. A \textbf{71}, 044301 (2005).
\bibitem{xldp04} Li Xiao, Gui Lu Long, Fu-Guo Deng, and Jian-Wei Pan, Phys. Rev. A
\textbf{69}, 052307 (2004)
\bibitem{beige} A. Beige, B.-G. Englert, Ch. Kurtsiefer, and
H. Weinfurter, Acta Phys. Pol. A \textbf{101}, 357 (2002).
\bibitem{Bostrom} K. Bostr\"{o}em and T. Felbinger, Phys. Rev. Lett. \textbf{89}, 187902 (2002).
\bibitem{Deng} F. G. Deng, G. L. Long, and X. S. Liu, Phys. Rev. A \textbf{68}, 042317 (2003).
\bibitem{denglong} F. G. Deng and G. L. Long, Phys. Rev. A \textbf{69}, 052319 (2004).
\bibitem{cai1} Q. Y. Cai and B. W. Li, Chin. Phys. Lett. \textbf{21}, 601 (2004).
\bibitem{cai4} Q. Y. Cai and B. W. Li, Phys. Rew. A \textbf{69}, 054301 (2004).
\bibitem{jwang1} J. Wang, Q. Zhang and C. J. Tang, quant-ph/0511092.
\bibitem{jwang2} J. Wang, Q. Zhang and C. J. Tang, quant-ph/0602166.
\bibitem{cw1} C. Wang, F. G. Deng, Y. S. Li, X. S. Liu and G. L. Long, Phys. Rev. A \textbf{71}, 044305 (2005).
\bibitem{cw2} C. Wang, F. G. Deng and G. L. Long, Opt. Commun. \textbf{253}, 15 (2005).
\bibitem{tg} T. Gao, F. L. Yan and Z. X. Wang, quant-ph/0406083.
\bibitem{zjz} Z. J. Zhang and Z. X. Man, quant-ph/040321.
\bibitem{cai2} Q. Y. Cai, Phys. Rew. Lett. \textbf{91}, 109801 (2003).
\bibitem{cai3} Q. Y. Cai, Phys. Lett. A \textbf{351}, 23 (2006).
\bibitem{deng-core} F. G. Deng and G. L. Long, Phys. Rev. A \textbf{68}, 042315 (2003).
\bibitem{zhu} A. D. Zhu, Y. Xia, Q. B. Fan and S. Zhang, Phys. Rev. A \textbf{73}, 022338 (2006).
\bibitem{long} G. L. Long and X. S. Liu, Phys. Rev. A \textbf{65}, 032302 (2002).







\end{thebibliography}
\end{document}